# Evaluation of the phi-3-mini SLM for identification of texts related to medicine, health, and sports injuries

Chris Brogly
Department of Computer Science
Lakehead University
Orillia, Canada
cbrogly@lakeheadu.ca

Saif Rjaibi
Injury Prevention Research Office
St. Michael's Hospital
Toronto, Canada
Saif.Rjaibi@UnityHealth.to

Charlotte Liang
Injury Prevention Research Office
St. Michael's Hospital
Toronto, Canada
Charlotte.Liang@mail.utoronto.ca

Erica Lam
Injury Prevention Research Office
St. Michael's Hospital
Toronto, Canada
Erica.Lam@UnityHealth.to

Edward Wang
Injury Prevention Research Office
St. Michael's Hospital
Toronto, Canada
Edward.Wang@UnityHealth.to

Sarah Paleczny
Injury Prevention Research Office
St. Michael's Hospital
Toronto, Canada
Sarah.Paleczny@UnityHealth.to

Adam Levitan
Injury Prevention Research Office
St. Michael's Hospital
Toronto, Canada
Adam.Levitan@UnityHealth.to

Michael D. Cusimano
Injury Prevention Research Office
St. Michael's Hospital
Toronto, Canada
injury.prevention@UnityHealth.to

*Abstract*— Small Language Models (SLMs) have potential to be used for automatically labelling and identifying aspects of text data for medicine/health-related purposes from documents and the web. As their resource requirements are significantly lower than Large Language Models (LLMs), these can be deployed at the edge on several devices. SLMs often are benchmarked on health/medicine-related tasks, such as MedQA, although performance on these can vary especially depending on the size of the model in terms of number of parameters. Furthermore, these test results may not necessarily reflect real-world performance regarding the automatic labelling or identification of texts in documents and the web. As a result, we compared topic-relatedness scores from Microsoft's phi-3-mini SLM to the topic-relatedness scores from 7 human evaluators on 1144 samples of medical/health-related texts and 1117 samples of sports injury-related texts. These texts were from a larger dataset of about 9 million news headlines, each of which were processed and assigned scores by phi-3-mini. Our sample was selected (filtered) based on 1 (low filtering) or more (high filtering) Boolean conditions on the assigned scores. We found low-moderate correlation between the scores from the SLM and human evaluators for sports injury texts with low filtering ($\rho = 0.3413$) and medicine/health texts with high filtering ($\rho = 0.3854$), and low correlation for medicine/health texts with low filtering ($\rho = 0.2255$) and finally negligible correlation for sports injury-related texts with high filtering ($\rho = 0.0318$). (*Abstract*)

*Keywords— small language models, generative AI, model evaluation, human evaluation* (key words)

## I. INTRODUCTION

Identifying or labelling specific items related to health/medicine in the large amount of text from documents in clinical settings or broadly on the web is challenging. Traditional machine learning models are trained to predict for specific types of texts only. Sequence-to-sequence models based on BERT have been used successfully to extract specific types of text from documents although these end up being trained for specific instances and are not always quickly generalizable requiring fine-tuning. Large language models (LLMs) can encode clinical knowledge [1] and are being used for a variety of tasks in biomedical settings [2]. Performance benchmarks for LLMs on tasks such as MedQA demonstrate capabilities for medicine/health [3]. LLMs also perform well on benchmarks in other areas [4]. Given their capabilities with natural language understanding (NLU) tasks, these models are logical candidates for processing a wide variety of previously unseen text data with no additional startup effort.

LLMs publicly are typically cloud and GPU-based (ChatGPT, Bing). While this results in wide availability and good performance, there are some potential downsides to using LLMs to process large amounts of new text data with these, especially with regards to medicine/health. First of all, access is provided by different companies; analyzing sensitive data on the cloud can be problematic especially if inputs are being monitored; there is also network transfer of data. Secondly, cloud compute costs on large datasets may escalate. As a result, under some conditions exploring offline options may be beneficial and also possibly more cost-effective.

Small Language Models (SLMs) like phi-3 have similar capabilities to LLMs although have reduced parameters and generally reduced performance on NLU tasks [3]. The advantage of this class of generative models though is that they have lower hardware requirements and potentially can run on a number of different edge devices, fully offline if needed. Some versions of different SLMs like phi-3 can even run on smartphones, although at a reduced token generation rate given the lower hardware speed [3].

We were interested in exploring the possibility of using SLMs to automatically classify millions of sentence-length texts related to topics in health/medicine. In part, this was due to access to various offline health-related text data, but also that we were interested in connecting an existing dataset of 9 million+ Canadian news website headlines/links to different sources of health data to determine if media mentions might be related to these topics. A previous analysis from our research lab analyzed North American newspapers for trends in brain injuries using qualitative techniques [5]; applying an SLM to classify texts for a quantitative analysis with a large sample size would only require existing compute resources for us and no manual intervention. We anticipated this would work based on some early observations using the phi-3-mini SLM, so we processed the existing 9 million+ Canadian news headline/link dataset through the phi-3-mini SLM by asking it to assign relatedness scores of 1-10 for 7 medical/health-related topics for every headline/link text. Although we suspected that performance on web data would be acceptable, it remained unclear just how effective an SLM would be in determining the relatedness to some of these topics in comparison to human evaluators. Additionally, it was not clear if the SLM performed well just by assigning one topic-relatedness score to a text or if conditions taking into account multiple scores would improve the utility of the SLM predictions overall, or, in other words, if conditions would filter out bad/erroneous predictions from the SLM. As a result, we used 7 human evaluators with degrees in health sciences to evaluate a sample of 1144 predictions for medical/health-related texts and 1117 predictions for sports injury-related texts taken from our 9-million+ headline/link text database, to compare them to the ratings from phi-3-mini.

## II. METHODS

### A. Medical/health-related topics chosen

Initially, each of the 9 million+ headlines/links in our news dataset was rated by phi-3-mini. The model was provided with a context consisting of "You are helping to determine if text is related to T" where T was the relatedness topic, and a question consisting of "On a scale of 1-10, is the following about T? Report only the number: ". The model usually would report only one numeric value as specified but on rare occasions would output a text explanation; we did not consider those explanations here. Example inputs to the model along with the list of chosen topics, including those that were selected for this evaluation are listed in Table 1 below. Eventually, only medicine/health and sports injury-related texts were selected for this evaluation, and samples of 1144 and 1117 texts were taken for each of those categories, respectively.

TABLE 1: OVERVIEW OF SLM INPUTS

| Context | "You are helping to determine if text is related to (T)" |
|---|---|
| Question | "On a scale of 1-10, is the following about (T)? Report only the number: " |
| Full list of topics analyzed by phi-3-mini | Medicine/health, cannabis, sports injuries, opioids, firearm injury, traffic accidents, hospitalization, sentiment |
| Topics (T) considered in this work | Medicine/health, sports injuries |
| Medical/health sample input | "Repeat floods prompt health concerns over persistent mould" |
| Sports injury sample input | "Habs G Allen out day-to-day with upper-body injury" |

### B. Dataset processing

The Canadian news headline/link dataset consists of 9,353,430 texts from the Common Crawl over January 3rd, 2017-June 27th 2023, with data from every Tuesday and Friday. This dataset was initially processed to study clickbait levels on Canadian news sites but is repurposed here to determine medical/health and sports injury topic relatedness (among other categories that are not considered here, listed in Table 1), where each headline/link was rated on a scale of 1-10 for the category by the phi-3-mini SLM. We selected this Canadian news text dataset as it was a substantial size although still small enough to be processed in a reasonable amount of time by several enthusiast-grade desktop GPUs. We processed the dataset in parallel with 1 instance of the CUDA phi-3-mini model running on 2x NVIDIA GeForce GTX 1080Ti cards and 2 instances of the phi-3-mini model running on 1x NVIDIA GeForce RTX 4090. We were able to process about 475,000 texts per week on each 1080Ti and 950,000 texts per week, per model on the 4090. Our software was not optimized for performance or job management due to time/programming resource constraints.

### C. Human evaluation of SLM topic-relatedness scores

For our total sample of 2261 texts, there were 7 human evaluators of phi-3-mini predictions, and each held degrees in health-related fields and were actively contributing to research projects either as graduate students, research assistants, or research visitors at the Injury Prevention Research Office at St. Michael's Hospital in Toronto, Canada. First, all seven raters met to benchmark the rating of 20 medical/health-related headlines and 20 sports injury-related headlines randomly selected from those processed by phi-3-mini. Evaluators were asked to arrive at a consensus on (i) a binary yes/no evaluation for topic-relatedness, and (ii) a numerical 1-10 rating of how related the text was to the topic. Next, all evaluators were provided with an identical set of 40 medical/health-related headlines and 40 sports injury-related headlines randomly selected from the list of all headlines evaluated by the SLM, and were asked to independently provide both a binary and numerical 1-10 rating. The inter-rater agreement was assessed through Fleiss's Kappa of the binary ratings, and pairwise correlation and intraclass correlation coefficients (ICC; two-way random effects) of the numerical ratings; these are reported on in results.

Once interrater agreement was established, evaluators were separately provided with 200 random phi-3-mini processed texts (with the phi-3-mini topic ratings withheld), for each of 4 categories; 1) medicine/health, low filtering, 2) medicine/health, high filtering, 3) sports injuries, low filtering, 4) sports injuries, high filtering. Raters were asked to complete at least 50 from each, and to provide (i) a binary yes/no evaluation, and (ii) a numerical 1-10 rating. The idea with the "filtering" was that it was initially observed with the phi-3-mini model that most of the unfiltered text it analyzed from our web news data was completely unrelated to our categories, but many of these were being given scores greater than 1 which was incorrect. Many less errors were observed from the model based on inspection when we looked at only the higher rated scores, such as those > 7, although errors would still be found. So, based on this observation, for any useful performance comparison to occur, the strategy was just to provide only texts that had a rating > 7 on the 1-10 scale to our human raters, otherwise the disagreement between the raters and the model would have always been consistently very high. As a result, our low-filtering datasets consist of texts that the phi-3-mini model rated > 7 for the topic-relatedness and our high-filter samples include this condition and also additional conditions, using some of the other topic category relatedness scores, to see if any agreement between raters and the model would improve. For health/medicine additional conditions include relatedness scores for cannabis, sports injuries, and opioids being equal to 1, and for sports injuries health/medicine being > 7 and opioids and cannabis being equal to 1. Filtering essentially is applying these necessary conditions given the amount of errors when using unfiltered data.

## III. RESULTS

Additional information, such as sample sizes, about our Canadian news headline/link dataset and the interrater task along with the final rating task are seen in Table 2. As previously stated, seven human evaluators were provided an identical set of 40 medical/health-related headlines and 40 sports injury-related headlines randomly selected from the list of all headlines evaluated by the SLM, and asked to independently provide both a binary yes/no and numerical 1-10 rating of how related the headline was to the topic. For the numerical ratings, strong interrater correlations are shown in Table 3 and Table 4. Table 5 displays other interrater results as follows: ICC was 0.758 for medical/health-related and 0.640 for sports injury-related headlines. Inter-rater pairwise correlation coefficients (PCCs) ranged from 0.729 to 0.919, with an average of 0.798 and standard deviation of 0.053, and ranged from 0.426 to 0.943, with an average of 0.696 and a standard deviation of 0.160 for sports injury. For the binary ratings, we omit correlation tables for brevity as this was a secondary job for human evaluators. Table 6 displays results relevant to this task as follows: Fleiss's Kappa was 0.789 for medical/health-related and 0.648 for sports injury-related headlines. PCCs ranged from 0.657 to 0.933, with an average of 0.800 and standard deviation of 0.078 for medical/health, and ranged from 0.293 to 0.901, with an average of 0.638 and standard deviation of 0.176 for sports injury.

TABLE 2: PROCESSED NEWS DATASET STATISTICS

| Dataset Property | Value |
| --- | --- |
| Total number of texts | 9,353,430 |
| HTML tags processed | "a", "span", "h1", "h2", "h3", "h4", "h5" |
| Domains total | 236 |
| Total number of texts evaluated by phi-3-mini and human evaluators | 2261 |
| Initial interrater evaluation task size, medical/health | 40 texts |
| Initial interrater evaluation task size, sports injuries | 40 texts |
| Final evaluation, medicine/health, low filter | 50 min, 200 max texts |
| Final evaluation, medicine/health, high filter | 50 min, 200 max texts |
| Final evaluation, sports injuries, low filter | 50 min, 200 max texts |
| Final evaluation, sports injuries, high filter | 50 min, 200 max texts |

TABLE 3: PAIRWISE RATER CORRELATION COEFFICIENTS FOR NUMERIC RATINGS OF 40 MEDICAL/HEALTH-RELATED TEXTS.

|  | Rater 1 | Rater 2 | Rater 3 | Rater 4 | Rater 5 | Rater 6 | Rater 7 |
| --- | --- | --- | --- | --- | --- | --- | --- |
| Rater 1 | 1 | 0.9193 | 0.7622 | 0.7549 | 0.7901 | 0.8123 | 0.7954 |
| Rater 2 | 0.9193 | 1 | 0.7518 | 0.7882 | 0.7742 | 0.8907 | 0.7474 |
| Rater 3 | 0.7622 | 0.7518 | 1 | 0.7345 | 0.7737 | 0.8341 | 0.7290 |
| Rater 4 | 0.7549 | 0.7882 | 0.7345 | 1 | 0.8875 | 0.8068 | 0.8415 |
| Rater 5 | 0.7901 | 0.7742 | 0.7737 | 0.8875 | 1 | 0.7992 | 0.8209 |
| Rater 6 | 0.8123 | 0.8907 | 0.8341 | 0.8068 | 0.7992 | 1 | 0.7490 |
| Rater 7 | 0.7954 | 0.7474 | 0.7290 | 0.8415 | 0.8209 | 0.7490 | 1 |

TABLE 4: PAIRWISE RATER CORRELATION COEFFICIENTS FOR NUMERIC RATINGS OF 40 SPORTS INJURY-RELATED HEADLINES.

|  | Rater 1 | Rater 2 | Rater 3 | Rater 4 | Rater 5 | Rater 6 | Rater 7 |
| --- | --- | --- | --- | --- | --- | --- | --- |
| Rater 1 | 1 | 0.5839 | 0.7929 | 0.5763 | 0.6255 | 0.9428 | 0.8346 |
| Rater 2 | 0.5839 | 1 | 0.7679 | 0.9066 | 0.9403 | 0.5595 | 0.4385 |
| Rater 3 | 0.7929 | 0.7679 | 1 | 0.7717 | 0.7663 | 0.7425 | 0.6436 |
| Rater 4 | 0.5763 | 0.9066 | 0.7717 | 1 | 0.8552 | 0.5732 | 0.4256 |
| Rater 5 | 0.6255 | 0.9403 | 0.7663 | 0.8552 | 1 | 0.5635 | 0.5035 |
| Rater 6 | 0.9428 | 0.5595 | 0.7425 | 0.5732 | 0.5635 | 1 | 0.7936 |
| Rater 7 | 0.8346 | 0.4385 | 0.6436 | 0.4256 | 0.5035 | 0.7936 | 1 |

TABLE 5: AGREEMENT BETWEEN SEVEN HUMAN EVALUATORS IN THEIR NUMERIC RATINGS.

|  |  | Medical/Health | Sports Injury |
|---|---|---|---|
| Numerical ratings | ICC | 0.7583 | 0.6399 |
|  | Min PCC | 0.7291 | 0.4256 |
|  | Max PCC | 0.9194 | 0.9428 |
|  | Average PCC | 0.7983 | 0.6956 |
|  | Standard Dev. PCC | 0.0528 | 0.1595 |

TABLE 6: AGREEMENT BETWEEN SEVEN HUMAN EVALUATORS IN THEIR BINARY RATINGS.

|  |  | Medical/Health | Sports Injury |
|---|---|---|---|
| Binary ratings | Fleiss's Kappa | 0.7893 | 0.6476 |
|  | Min PCC | 0.6567 | 0.2933 |
|  | Max PCC | 0.9333 | 0.9010 |
|  | Average PCC | 0.7997 | 0.6380 |
|  | Standard Dev. PCC | 0.0781 | 0.1760 |

TABLE 7: AGREEMENT BETWEEN SEVEN HUMAN EVALUATORS AND THE PHI-3 SLM IN BINARY (YES/NO) AND NUMERIC RATINGS (1-10) FOR MEDICAL/HEALTH (LOW/HIGH FILTER) AND SPORTS INJURY (LOW/HIGH FILTER)-RELATED HEADLINES.

|  |  | Medical/health texts, low filtering | | Medical/health texts, high filtering | | Sports-injury texts, low filtering | | Sports-injury texts, high filtering | |
|---|---|---|---|---|---|---|---|---|---|
| Headlines Evaluated |  | 609 | | 535 | | 538 | | 579 | |
| Binary ratings |  |  | | | | | | | |
|  | % Agreement | 0.5468 | | 0.7458 | | 0.0669 | | 0.2401 | |
|  |  |  | | | | | | | |
| Numerical ratings |  | Human Evaluators | Phi-3 | Human Evaluators | Phi-3 | Human Evaluators | Phi-3 | Human Evaluators | Phi-3 |
|  | Mean | 5.2475 | 9.5776 | 6.3178 | 9.3682 | 1.8662 | 9.8717 | 3.3148 | 9.8174 |
|  | Median | 6.0000 | 10.000 | 7.0000 | 10.000 | 1.0000 | 10.000 | 1.0000 | 10.000 |
|  | Mode | 1.0000 | 10.000 | 7.0000 | 10.000 | 1.0000 | 10.000 | 1.0000 | 10.000 |
|  | Standard Dev. | 3.5608 | 0.7734 | 2.9452 | 0.8810 | 2.0159 | 0.4847 | 3.4206 | 0.5535 |
|  | Spearman corr. | 0.2255 | | 0.3854 | | 0.3413 | | 0.0318 | |

The results for the full evaluation task are shown above in Table 7. With respect to the binary (yes/no) ratings, the phi-3-mini model rated all of the texts provided to evaluators as > 7, which means that the model stated they were related to one of the two topics. As a result, the question became whether or not a human evaluator agreed with that decision. For medical/health texts, depending on filtering, there was mixed agreement from the human evaluators (low filtering, 54%), to good agreement (high filtering, 74%). For sports-injury texts, depending on filtering, there was negligible (low filtering, 6%) to marginal (high filtering, 24%) agreement that the texts were in fact related to the topic.

With respect to the numeric ratings, based on Spearman correlations, in medical/health-related texts there was low ($\rho = 0.23$, low filtering) to low-moderate ($\rho = 0.38$, high filtering) correlation between phi-3-mini ratings and the human evaluators. In sports injury-related texts there was low-moderate ($\rho = 0.34$, low filtering) to negligible correlation ($\rho = 0.03$, high filtering) between phi-3-mini ratings and the human ratings. We speculate on why binary agreement is acceptable for medical/health texts but low for sports injuries, and why there is some consistent numeric correlation for medical/health texts, but not sports injuries, in the discussion section.

IV. DISCUSSION

With respect to binary ratings, it is suspected that there is better identification of medical/health texts in comparison to sports injury texts because the SLM most likely did not have specific sports injury training data. However, it was benchmarked on MedQA as mentioned in the introduction, which might explain some of the better performance for medical/health. Additionally, it is possible that there needed to be stricter filters on sports injury than what ended up being used in this study, as for most of the texts that the SLM scored as > 7, the human evaluators disagreed that they were even related to sports injuries at all. The results for numeric rating correlations are somewhat comparable. For medical/health ratings, they appear similar to the binary ratings, although the correlation coefficients range from low ($\rho = 0.22$) to low-moderate ($\rho = 0.38$) and seem weaker overall than our binary result. For sports injuries, there is a low-moderate ($\rho = 0.34$) correlation with human evaluators when using low filtering (sports injury relatedness > 7, no additional conditions) but a negligible result for high filtering (which includes medical/health-relatedness > 7 and cannabis/opioid-

relatedness = 1). It is suspected the negligible numeric result occurred because human evaluators disagreed more with the extent how these texts were related.

Based on the results, we suspect that human evaluators at the time of writing are more effective at directly answering a question like "On a scale of 1-10, is the following about sports injuries?". This however comes with some trade-offs. Even though manual evaluation may be more accurate, it is not feasible for evaluators to provide ratings for millions of texts from the web or other sources, but SLMs can. Additionally, it became evident the SLM had unique capabilities when processing this text. For instance, we noticed that when a text like "Reporter_Name, The Associated Press" (which at first glance may appear to be unrelated junk text) was passed into the SLM, it would at times rate these texts as being highly related to medicine/health or especially sports injuries which seemed unusual. When we went and checked which journalist's name was part of this text, it turned out they actually were sports or health reporters. We suspect that similar understandings may be happening with other aspects of text as well but further analysis is needed to confirm. In general however, there had to be some type of boolean filtering on the results because if the $> 7$ condition was not applied, the SLM had a high misclassification rate on initial inspection and our human evaluation results would have been far lower than what they currently are.

## V. LIMITATIONS

While SLMs have great potential to be deployed on edge devices, using them is still, in general, slow on very large datasets, even with top-of-the-line consumer-grade GPUs at the time of writing like the RTX 4090. It is not feasible to process any size of data unless the GPUs are scaled out, and the dataset used here with 2 1080Tis and 1 4090 took about 4 weeks. If American or worldwide news outlets had been considered for this analysis instead of the Canadian news dataset used here, it would not have been possible to process those in a reasonable amount of time. Still, an SLM is much faster than human evaluators – it would take years to manually evaluate 9 million headlines, compared to 4 weeks with an SLM and good hardware. However, there are likely more accuracy issues with the SLM. For the evaluators, due to time constraints, there had to be a limit on the number of texts that they examined manually and so 50 per category was selected for each evaluator. Since there were only so many evaluators in this analysis, the total texts we were able to examine is 2261, so there may be limitations in our results due to the sample size.

## VI. CONCLUSION

Medical/health and sports injury topic-relatedness ratings from the phi-3-mini SLM were evaluated based on the judgements of 7 human raters. To the best of our knowledge, phi-3 and comparable SLMs are not generally evaluated in a similar way and as such this paper contributes to the real-world performance discussion of phi-3-mini and SLMs in general.

ACKNOWLEDGEMENT

This work was supported by a grant from Lakehead University, Faculty of Science and Environmental Studies. This work was supported by the Injury Prevention Research Office at St. Michael's Hospital.

REFERENCES

[1] K. Singhal *et al.*, "Large language models encode clinical knowledge," *Nature*, 2023, doi: 10.1038/s41586-023-06291-2.

[2] A. J. Thirunavukarasu, D. S. J. Ting, K. Elangovan, L. Gutierrez, T. F. Tan, and D. S. W. Ting, "Large language models in medicine," 2023. doi: 10.1038/s41591-023-02448-8.

[3] Microsoft, "Phi-3 Technical Report: A Highly Capable Language Model Locally on Your Phone," 2024. [Online]. Available: https://arxiv.org/pdf/2404.14219?

[4] Y. Chang *et al.*, "A Survey on Evaluation of Large Language Models, (Accepted by ACM Transactions on Intelligent Systems and Technology (TIST))," 2023.

[5] M. D. Cusimano, B. Sharma, D. W. Lawrence, G. Ilie, S. Silverberg, and R. Jones, "Trends in North American Newspaper Reporting of Brain Injury in Ice Hockey," *PLoS One*, 2013, doi: 10.1371/journal.pone.0061865.